\documentclass[aps,twocolumn,prl,
groupedaddress,nofootinbib,nobalancelastpage,nobibnotes]{revtex4}
\pdfoutput=1
\usepackage{amsmath,amsfonts,amssymb,mathrsfs,graphicx}
\usepackage[squaren]{SIunits}

\newcommand{\eg}{{\it e.g.}}

\newcommand{\eq}{Eq.}

\newcommand{\fig}{Fig.}

\newcommand{\Ref}{Ref.}
\newcommand{\Refs}{Refs.}

\newcommand{\equ}[1]{\eq~(\ref{equ:#1})}
\newcommand{\figu}[1]{\fig~\ref{fig:#1}}

\newcommand{\bi}{\begin{itemize}}
\newcommand{\ei}{\end{itemize}}

\begin{document}

\title{Magnetic Field and Flavor Effects on the Gamma-Ray Burst Neutrino Flux}

\author{Philipp Baerwald}
\affiliation{Institut f{\"u}r theoretische Physik und
  Astrophysik, Universit{\"a}t W{\"u}rzburg, Am Hubland, D-97074 W{\"u}rzburg, Germany}

\author{Svenja H{\"u}mmer}
\affiliation{Institut f{\"u}r theoretische Physik und
  Astrophysik, Universit{\"a}t W{\"u}rzburg, Am Hubland, D-97074 W{\"u}rzburg, Germany}

\author{Walter Winter}
\affiliation{Institut f{\"u}r theoretische Physik und
  Astrophysik, Universit{\"a}t W{\"u}rzburg, Am Hubland, D-97074 W{\"u}rzburg, Germany}

\date{\today}

\begin{abstract}
We reanalyze the prompt muon neutrino flux from gamma-ray bursts (GRBs), at the example of the often used reference Waxman-Bahcall GRB flux, in terms of the particle physics involved. We first reproduce this reference flux treating synchrotron energy losses of the secondary pions explicitly. Then we include additional neutrino production modes, the neutrinos from muon decays, the magnetic field effects on all secondary species, and flavor mixing with the current parameter uncertainties. 
 We demonstrate that the combination of these effects modifies the shape of the original Waxman-Bahcall GRB flux significantly, and changes the normalization by a factor of three to four.  As a consequence, the gamma-ray burst search strategy of neutrino telescopes may be based on the wrong flux shape, and the constraints derived for the GRB neutrino flux, such as the baryonic loading, may in fact be already much stronger than anticipated.
\end{abstract}


\maketitle

%
Neutrino telescopes, such as IceCube~\cite{Ahrens:2003ix} or ANTARES~\cite{Aslanides:1999vq},  are designed to detect neutrinos from astrophysical sources. There are numerous candidate sources, see \Ref~\cite{Becker:2007sv} for a review and \Ref~\cite{Rachen:1998fd} for the general theory. We focus on the prompt emission of gamma-ray bursts (GRBs) in this letter, where photohadronic interactions are expected to lead to a significant flux of neutrinos~\cite{Waxman:1997ti}. So far, no extraterrestrial high energy neutrino flux has been detected yet. That is, for sources optically thin to neutrons, consistent with generic bounds~\cite{Waxman:1998yy,Mannheim:1998wp} which are just being touched by IceCube. The search for GRB neutrinos has been driven by analytical estimates for the shape and normalization, the simplest one being the Waxman-Bahcall (WB) flux~\cite{Waxman:1998yy}. More recent analyses, such as the stacking analysis in \Ref~\cite{Abbasi:2009ig}, relating the neutrino flux to the observed gamma-ray flux,  are based on the analytical generalization of this flux for arbitrary input parameters following \Ref~\cite{Guetta:2003wi}. These calculations typically approximate the $\Delta(1232)$ resonance for the charged pion production
\begin{equation}
	p + \gamma \rightarrow \Delta^+ \rightarrow \left\{\begin{array}{lc} n + \pi^+ & \text{1/3 of all cases} \\ p + \pi^0 & \text{2/3 of all cases} \end{array} \right. \label{equ:Delta}
\end{equation}
in some form.
However, the GRB neutrino flux computation  has been updated over the last ten years from the particle physics point of view by improving the description of the photo-meson production processes, and it has been  obvious there is a substantial impact from magnetic field effects and flavor mixing on the neutrino flux as well; see, \eg,  \Refs~\cite{Mucke:1999yb,
Murase:2005hy,Kashti:2005qa,Lipari:2007su,Hummer:2010vx}. 
In this letter, we make the impact of these effects very explicit by revising the often used WB reference flux from \Ref~\cite{Waxman:1998yy}. We include the relevant pion production modes and neutrinos from kaon and neutron decays. We treat the magnetic field effects on each charged particle species explicitly, and we include flavor effects/flavor mixing. Note that we keep our considerations as independent of the astrophysical source model as possible to factor out the particle physics effects, which are much better known than the details of the astrophysical model. The purpose of this letter is to demonstrate how the original WB flux changes in both shape and normalization effect by effect, and where the main impact comes from. We also discuss the impact on data analyses. The technology used in this letter is based on \Refs~\cite{Hummer:2010vx,Hummer:2010ai}, where details can be found.

\begin{figure}[t]
	\centering
	\includegraphics[width=0.85\columnwidth]{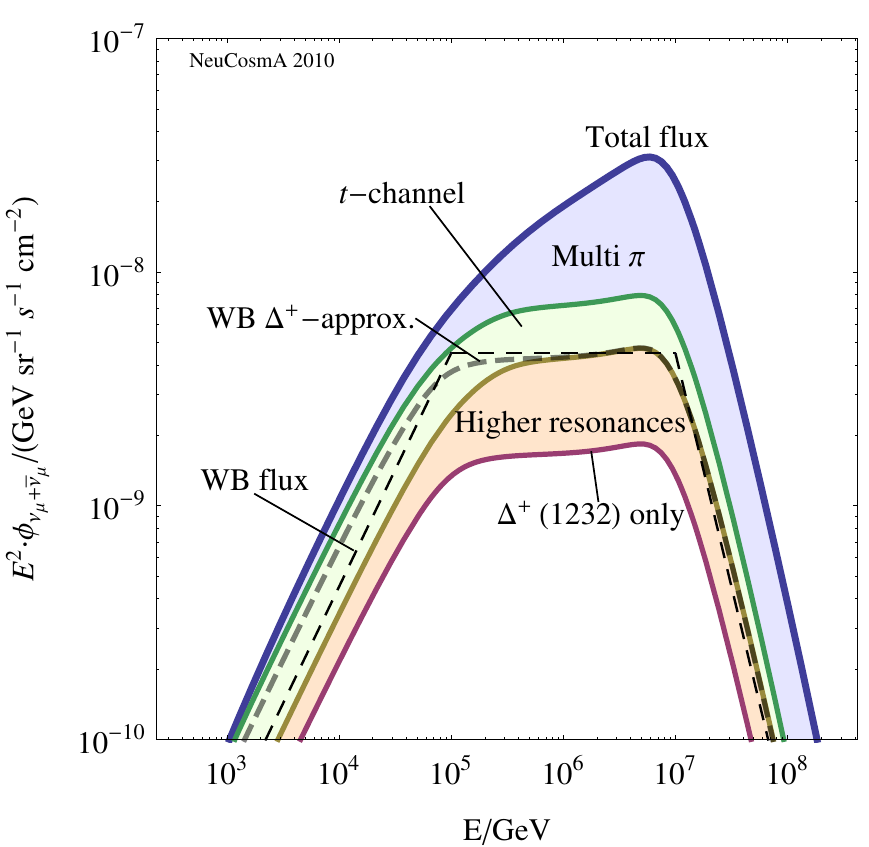}
	\caption{\label{fig:photo} The WB flux from \equ{WB} (thin dashed curve), the numerically reproduced flux using the $\Delta^+$ resonance only (lower solid curve), and the WB flux including higher resonances, direct production/$t$-channel processes, and multi pion production (high energy processes), which are successively switched on, leading to the final upper solid curve. Here the $\nu_\mu$ flux from $\pi^+$ and $\pi^-$ decays is considered. The normalization of our result to the numerically reproduced WB flux (gray dashed curve) is described in the main text.}
\end{figure}

%
In the standard picture, protons collide with photons, possibly from synchrotron emission of co-accelerated electrons or positrons (see, \eg, \Ref~\cite{Dermer:2003zv}), leading to pion production by processes as, for instance, \equ{Delta}. The charged pions then decay further into neutrinos, such as by $\pi^+  \rightarrow  \mu^+ + \nu_\mu$, $\mu^+ \rightarrow e^+ + \nu_e + \bar{\nu}_\mu$.
 For the shape of the WB flux, consider only the $\nu_\mu$ from pion decays for the moment. It is often assumed  that the target photon field corresponds to the observed prompt GRB flux, which is typically parameterized by 
$dN_{\gamma}(E)/dE \propto E^{\alpha_{\gamma}}$ for $E < \varepsilon_{\gamma,\text{break}}$ and  $dN_{\gamma}(E)/dE \propto E^{\beta_{\gamma}}$ for $E > \varepsilon_{\gamma,\text{break}}$ in the observer's frame, where $\alpha_\gamma \simeq -1$, $\beta_\gamma \simeq -2$, and the break $\varepsilon_{\gamma,\text{break}}$ at a few hundred keV.
If the protons are injected with a power law with injection index two, one obtains for the prompt GRB neutrino flux, referred to as ``WB flux'', 
\begin{equation}
	E^2_{\nu} \frac{\mathrm{d}N_{\nu}}{\mathrm{d}E_{\nu}} \propto \left\{ \begin{array}{ll} (E_{\nu} / \varepsilon^b_{\nu})^{\alpha_{\nu}} & \text{for} \; E_{\nu} < \varepsilon^b_{\nu} \\ (E_{\nu} / \varepsilon^b_{\nu})^{\beta_{\nu}} & \text{for} \; \varepsilon^b_{\nu} \leq E_{\nu} < \varepsilon^s_{\nu} \\ (E_{\nu} / \varepsilon^b_{\nu})^{\beta_{\nu}} (E_{\nu} / \varepsilon^s_{\nu})^{-2} & \text{for} \; E_{\nu} \geq \varepsilon^s_{\nu} \end{array} \right.  \label{equ:WB}
\end{equation}
with $\alpha_\nu = -\beta_\gamma - 1 \simeq +1$, $\beta_\nu = - \alpha_\gamma  - 1 \simeq 0$, $\varepsilon^b_{\nu} \simeq 10^5 \, \giga\electronvolt$ and $\varepsilon^s_{\nu} \simeq 10^7 \, \giga\electronvolt$. 
 For the analytical estimates of the break energies, we follow the treatment in \Ref~\cite{Guetta:2003wi}, assuming that $\Gamma = 10^{2.5}$ and $\textit{z} = 2$; see, \eg, \Refs~\cite{Guetta:2003wi,Wanderman:2009es}. 
The first break energy $\varepsilon^b_{\nu}$ can be related to  $\varepsilon_{\gamma,\text{break}}$  from the  threshold of the photohadronic interactions at the source. As a minor difference to \Ref~\cite{Guetta:2003wi}, where heads-on collisions between photons and protons are assumed for the threshold, we include the effect that the pion production efficiency peaks at higher center-of-mass energies (see Fig.~4 in \Ref~\cite{Hummer:2010vx}) to match our numerical results. This leads to a factor of two higher photon energy break ($14.8 \, \mathrm{keV}$) in the source frame 
to match the $\varepsilon^b_{\nu} \simeq 10^5 \, \giga\electronvolt$ for the chosen parameter set. The second break comes from pion cooling in the magnetic field. It can be computed from the energy where the pion decay rate equals the synchrotron loss rate. In order to reproduce $\varepsilon^s_{\nu} \simeq 10^7 \, \giga\electronvolt$, one has $B \simeq 3 \cdot 10^5 \, \mathrm{G}$. Note that, in the light of recent Fermi data, it is not clear how ``typical'' this parameter set is, which, however, does not affect the logic of this letter.
As another relevant parameter, we choose the maximum proton energy by balancing synchrotron loss and acceleration rates with an acceleration efficiency of 10\%~\cite{Hillas:1985is}. For the expected normalization of the flux in \equ{WB}, we use~\cite{Waxman:1998yy} (updated in \Ref~\cite{Waxman:2002wp})
\begin{equation}
E_\nu^2 \phi_\nu = 0.45 \cdot 10^{-8} \, \frac{f_\pi}{0.2} \, \giga\electronvolt \, \centi\meter^{-2} \, \second^{-1} \, \steradian^{-1}
\label{equ:wb}
\end{equation}
per neutrino species ($\nu_e$, $\nu_{\mu}$, or $\bar{\nu}_{\mu}$). After flavor mixing, the combined muon neutrino and antineutrino flux is, again, approximately given by \equ{wb}~\cite{Learned:1994wg}.
This estimate is based on the assumption that GRBs are a dominant cosmic ray source in which the high energy protons dissipate a  fraction $f_\pi < 1$ of energy into pion production before leaving the source. If no neutrino flux at the level of \equ{wb} is observed, it means that effectively the product of $f_\pi$ and the fraction of energy in protons (baryonic loading) becomes stronger constrained~\cite{Guetta:2003wi} -- and therefore the hypothesis of GRBs being the dominant cosmic ray source. We choose $f_\pi=0.2$ for the following figures.
For the numerical treatment of the photohadronic interactions, we follow \Ref~\cite{Hummer:2010vx} (Sim-B), based on the physics of SOPHIA~\cite{Mucke:1999yb}
 and the weak decays in \Ref~\cite{Lipari:2007su}, including the helicity dependence of the muon decays. The energy losses and other production modes are treated as described in \Ref~\cite{Hummer:2010ai}. 
We assume that synchrotron losses are the leading energy loss mechanism, which means that only the product of the proton and photon densities is required, see \eq~(7) of \Ref~\cite{Hummer:2010vx}. Therefore, our results are independent of the baryonic loading and GRB model details. We also assume that the source is optically thin to neutrons, which means that secondary interactions are neglected. There are limitations to these assumptions, such as if the protons cool significantly by photohadronic interactions (see, \eg, discussion in \Ref~\cite{Murase:2005hy}). However, such processes cannot be included in a model-independent way, because they require separate knowledge on the proton and photon densities.
\begin{figure*}[t]
	\centering
	\includegraphics[width=0.4\textwidth]{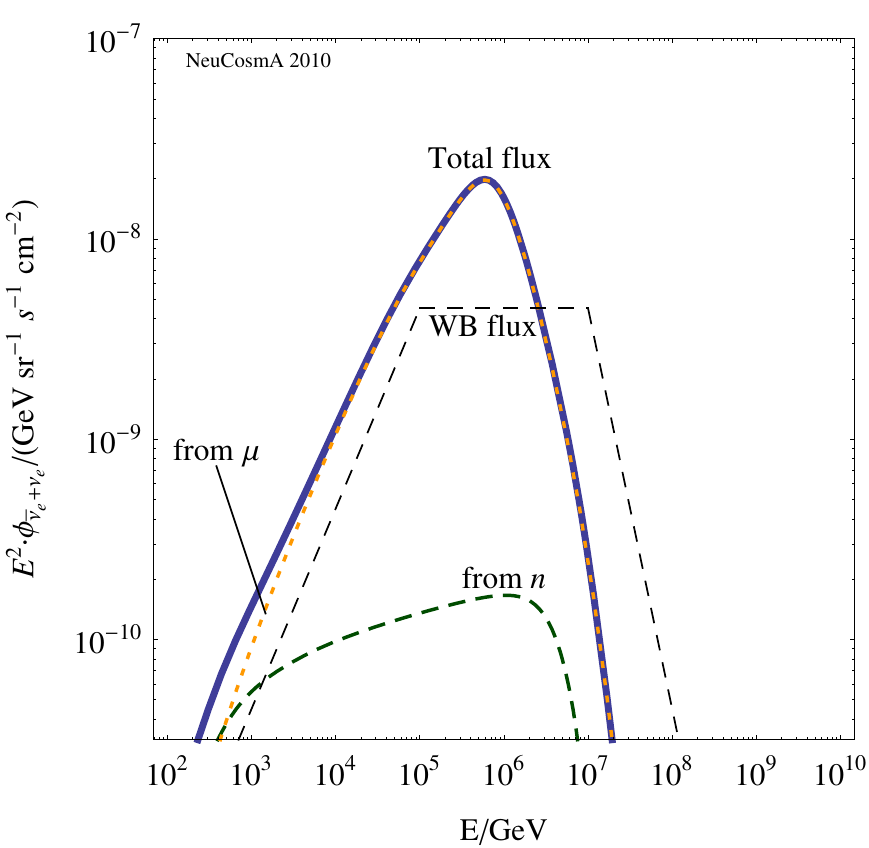} \hspace*{0.5cm}
	\includegraphics[width=0.4\textwidth]{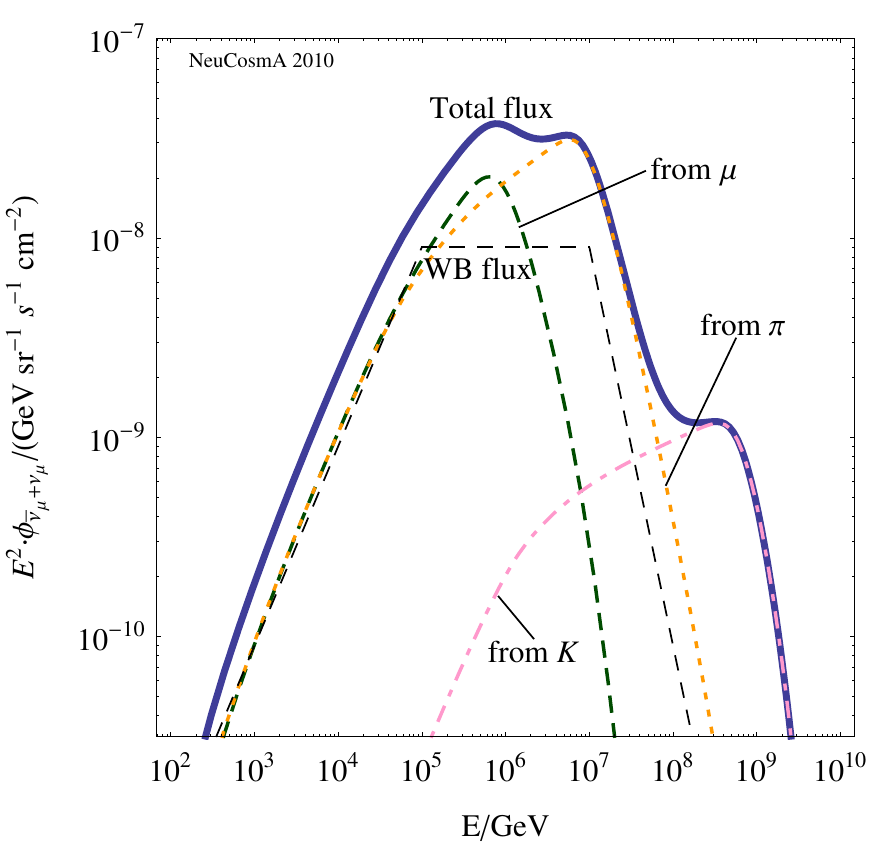}
	\caption{\label{fig:source} Total electron neutrino (left panel) and muon neutrino (right panel) flux before flavor mixing (thick solid curves), where the neutrino and antineutrino fluxes are added. The individual contributions to these neutrino fluxes from $\pi$, $\mu$, $n$, and $K$ decays are shown as well.  
}

\end{figure*}
We show in \figu{photo} the WB reference flux from \equ{WB} as thin dashed curve for $\nu_\mu$ from charged pion decays only.  In order to normalize our flux to this curve, we need to take into account that the assumptions for  the $\Delta$ resonance vary in the literature. Therefore, we first of all reproduce the WB flux numerically by including the synchrotron cooling of the pions explicitely, leading to the thick gray dashed curve ``WB $\Delta^+$-approx.''.  Here we use the same cross sections, pion multiplicities, and inelasticities as in \Ref~\cite{Waxman:1997ti}, and we choose the normalization such that the energy going into neutrinos is the same as for the analytical estimate. Note that for this curve, the second break is automatically reproduced by magnetic field effects and not put in by hand, which results in a small pile-up effect at the plateau. By this choice, the product of proton and photon density normalizations is fixed, and we can use the input spectra to compute the effects of the more refined interaction model. We show this by the solid curves, where higher resonances, direct ($t$-channel) production, and multi pion production are successively added to the actual $\Delta^+(1232)$ resonance process in \equ{Delta}; see, \eg, \Refs~\cite{Mucke:1999yb,Hummer:2010vx}. The final result exceeds the WB estimate by a factor of a few, especially at high energies.
The additional tilt of the spectrum comes from the multi pion cross section staying approximately constant for high interaction energies. In addition, note that all processes other than the actual $\Delta$ resonance include  $\pi^-$ production and two or multi pion production modes. From \figu{photo}, one can read off that the WB approximation basically includes the effect from direct production at low energies and higher resonances at high energies, but the high energy (or two and multi pion) contributions are clearly underestimated. In addition, one can read off that the effect of the additional production processes can lead to an up to one order of magnitude change of the flux, depending on the assumptions on the $\Delta$ resonance in the literature.

%
Apart from pion decays, neutrinos are produced from muon decays in the pion decay chain. In addition, kaons produced in the photohadronic interactions similar to pions may decay into neutrinos~\cite{Asano:2006zzb}. The main qualitative difference among charged pions, muons, and kaons are their different masses and lifetimes, leading to different energies of the second (synchrotron) break in \equ{WB}; see, \eg, Fig.~3 in \Ref~\cite{Hummer:2010ai}. This effect has interesting implications for the flavor ratio of the neutrinos, which changes as a function of energy~\cite{Kashti:2005qa}; see also \Ref~\cite{Kachelriess:2007tr}. Finally, any neutrino flux will be accompanied by a neutron flux, as it is obvious from \equ{Delta}. These neutrons are, however, not stable. If the neutrons do not interact, they will decay either within or outside the source by $n \rightarrow p + e^- + \bar\nu_e$, leading to cosmic rays and (inevitably) to an additional (almost coincident) $\bar\nu_e$ neutrino flux.  We show the total electron neutrino (left panel) and muon neutrino (right panel) flux before flavor mixing in \figu{source}, where the neutrino and antineutrino fluxes are added. The individual contributions to these neutrino fluxes from $\pi$, $\mu$, $n$, and $K^+$ decays (we only consider the leading kaon contribution mode) are shown as well. The WB flux from \equ{wb} is given as reference for the corresponding number of neutrino species. In the right panel (muon neutrinos), one can clearly see the hierarchy in the second break energy among neutrinos from $\mu$, $\pi$, and $K$ decays. In the left panel (electron neutrinos), the main contribution comes from muon decays. However, neutron decays show up at low energies. 

\begin{figure}[t]
	\centering
	\includegraphics[width=0.9\columnwidth]{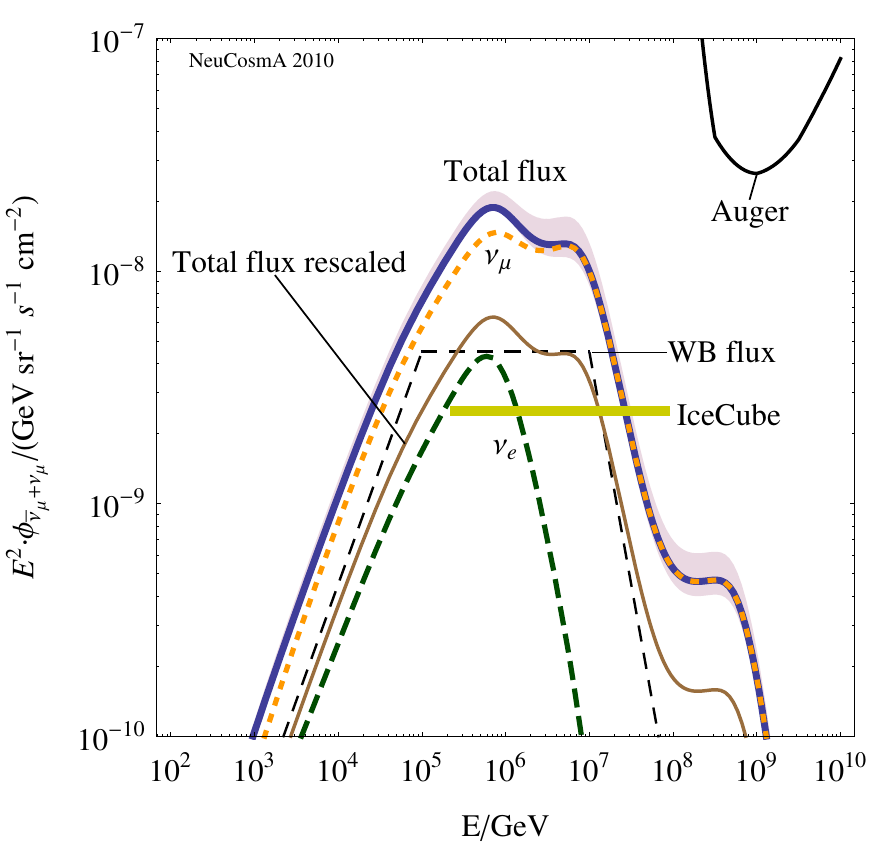}
	\caption{\label{fig:detector} Total muon neutrino flux after flavor mixing (dark thick solid curve). The individual contributions to this flux from 
muon and electron neutrinos are shown as well (from thick curves in \figu{source}). The WB flux from \equ{wb} is shown for reference, corrected by flavor mixing. In addition, a rescaled total flux is shown to illustrate the impact on the spectral shape. The shaded band shows the $3\sigma$ allowed range of the total flux from current mixing parameter uncertainties~\cite{Schwetz:2008er}. Here also the 10 year (extrapolated) full-scale limits from IceCube~\cite{Ahrens:2003ix} (for an $E^{-2}$ diffuse flux above the atmospheric neutrino background) and Auger~\cite{Abraham:2009uy} (differential limit) are estimated at the 90\% CL.
}
\end{figure}

%
In order to obtain the final muon neutrino flux relevant for muon tracks in neutrino telescopes, the total electron and muon neutrino fluxes in \figu{source} are superimposed by flavor mixing (averaged neutrino oscillations)~\cite{Learned:1994wg}, see \figu{detector}. The shaded band indicates the $3\sigma$ allowed range of the total flux from current mixing parameter uncertainties~\cite{Schwetz:2008er}. In fact, if the combined knowledge from the Double Chooz, Daya Bay, T2K, and NO$\nu$A is applied~\cite{Huber:2009cw}, as expected in about 2015, this band becomes hardly visible anymore. Therefore, mixing parameter uncertainties are less relevant for the GRB analysis, especially at the lower break (unless flavor ratios are considered). Comparing the final result (upper thick solid curve) with the WB flux (thin dashed curve), we notice that the expected neutrino flux is about a factor of three to four larger than the WB flux at 1~PeV.  In addition, we show a rescaled version of the final result (thin solid curve) to illustrate the impact on the spectral shape compared to the WB flux.  It is obvious from this comparison that the shape of \equ{WB} cannot be used for realistic data analyses or to search for point source GRBs. For instance, the first break, to which AMANDA and IceCube are most sensitive to, has basically disappeared in its original form. On the other hand, magnetic field and flavor effects lead to a characteristic double peak structure, one could search for if a few bursts dominated.  In addition, note the high energy excess coming from kaon decays, which increases the flux by at least one order of magnitude. At about $10^{8-9} \, \mathrm{GeV}$, horizontal air shower experiments, such as  Auger~\cite{Abraham:2009uy}, in fact have the best sensitivity, where the flux shown in \figu{detector} is representative for $\nu_\mu$ or $\nu_\tau$ events. Although the full-scale diffuse flux sensitivity is considerably above the expected neutrino flux, Auger may detect a flux from GRB kaon decays especially if a bright burst out-shines the cosmogenic neutrino flux. 
Therefore, neutrino point source studies should be initiated by these experiments.

%
In summary, we have revised the WB neutrino flux, often used as a reference GRB flux, by including the most relevant neutrino production processes, and by treating magnetic field and flavor effects explicitly. We have used as few assumptions as possible on the astrophysical source model.
We have demonstrated that the flux normalization increases by a factor of three to four with respect to the initial assumptions, and that the spectral shape exhibits a double peak structure qualitatively different from the WB flux. The main impact are additional neutrino production modes and magnetic field effects, which act differently on the charged secondary particle species. 
The revised spectral shape may allow for new search strategies for GRB neutrino fluxes. 
Since current state-of-the-art multi-messenger stacking analyses, such as \Ref~\cite{Abbasi:2009ig},  where the expected neutrino flux is computed from the observed gamma-ray flux on an event-by-event basis, rely on the assumptions on the photohadronic interactions, the non-observation of a flux may have stronger constraints on the energy equipartition between protons and electrons than anticipated. 

We would like to thank J. Becker, A. Kappes, and K. Murase for illuminating discussions and comments.


\end{document}